\documentstyle{article}
\def\rd{{\rm d}}
\def\rD{{\rm D}}
\def\be{\begin{equation}}
\def\ee{\end{equation}}
\def\bea{\begin{eqnarray}}
\def\eea{\end{eqnarray}}
\def\Pexp{{\rm Pexp}}
\def\rP{{\rm P}}
\begin{document}

\hfill{}

\vskip 3\baselineskip
\noindent{\Large\bf Nonlocal two--dimensional Yang--Mills- and generalized
Yang--Mills-theories}

\vskip \baselineskip

Khaled Saaidi$^{\; a,1}$ and Mohammad Khorrami$^{\; b,c,2}$

\vskip\baselineskip

\begin{itemize}
\item[{\small a:}]{\small Department of Physics, Tehran University,
North--Kargar Ave., Theran, Iran}
\item[{\small b:}] {\small Institute for Advanced Studies in Basic Sciences,
P. O. Box 159, Zanjan 45195, Iran}
\item[{\small c:}] {\small Institute for Studies in Theoretical Physics and
Mathematics, P. O. Box 5531, Theran 19395, Iran}
\item[{\small 1:}] {\small lkhled@molavi.ut.ac.ir}
\item[{\small 2:}] {\small mamwad@theory.ipm.ac.ir}
\end{itemize}

\vskip 2\baselineskip



\vskip 2\baselineskip

\begin{abstract}
A generalization of the two--dimensional Yang--Mills and generalized
Yang--Mills theory is introduced in which the building $B$--$F$ theory is
nonlocal in the auxiliary field. The classical and quantum properties of
this nonlocal generalization are investigated and it is shown that for large
gauge groups, there exist a simple correspondence between the properties
a nonlocal theory and its corresponding local theory.
\end{abstract}

\newpage
\section{Introduction}
Pure two--dimensional Yang--Mills theories (YM$_2$) have certain properties,
such as invariance under area--preserving diffeomorphisms and lack of
any propagating degrees of freedom. There are, however, ways to generalize
these theories without losing these properties. One way is the so--called
generalized Yang--Mills theories (gYM$_2$'s). In a YM$_2$, one starts
from a $B$--$F$ theory in which a Lagrangian of the form
$i{\rm tr}(BF)+{\rm tr}(B^2)$ is used. Here $F$ is the field--strength
corresponding to the gauge--field, and $B$ is an auxiliary field in the
adjoint representation of the gauge group. Carrying a path integral over
this field, leaves an effective Lagrangian for the gauge field of the form
${\rm tr}(F^2)$ \cite{BlTh}. In a gYM$_2$, on the other hand, one uses an
arbitrary class function of the auxiliary field $B$, instead of
${\rm tr}(B^2)$ \cite{Wit}. In \cite{GaSoYa} the partition function and the
expectation values of the Wilson loops for gYM$_2$'s were calculated. It is
worthy of mention that for gYM$_2$'s, one can not eliminate the auxiliary
field and obtain a Lagrangian for the gauge field. One can, however, use
standard path--integration and calculate the observables of the theory. This
was done in \cite{KhAl}.

To study the behaviour of these theories for large groups is also of
interest. This was studied in \cite{Rus} and \cite{DoKa} for ordinary
YM$_2$ theories and then in \cite{AgAlKh} for YM$_2$ and in \cite{AlKhAg}
and \cite{AlTo} for gYM$_2$ theories. It was shown that YM$_2$'s and
some classes of gYM$_2$'s have a third--order phase transition in a certain
critical area.

There is another way to generalize YM$_2$, and that is to use a non--local
action for the auxiliary field. An example of such theories is one in which
the $B^2$ action is replaced by an action
\be\label{1}
S_B:=w\left[\int\rd\mu\;{\rm tr}(B^2)\right],
\ee
or by
\be\label{2}
S_B:=w\left[\int\rd\mu\;\Lambda(B)\right],
\ee
where $\Lambda$ is a class function. We call these gerneralizations nonlocal
YM$_2$ (nlYM$_2$) and nonlocal gYM$_2$ (nlgYM$_2$), respectively. We want to
investigate the classical, and quantum behaviour of these theories, and also
their properties for large groups.

The scheme of the present paper is the following. In section 2, we introduce
the nlYM$_2$ and show that the classical solutions of YM$_2$ are also
classical solutions of nlYM$_2$.

In section 3, the wave functions and, as a special case, the partition
function of nlYM$_2$ on general surfaces are calculated.

In section 4, the properties of nlYM$_2$ large groups are investigated, and
it is shown that these properties can be related to those of YM$_2$ by
simple redefinitions.

Finally, in section 5 we introduce nlgYM$_2$ theories, and study their
properties, including the wave functions and large--group behaviour. It is
seen that the large--group properties of nlgYM$_2$ are easily obtained
from those of ordinary gYM$_2$, in a manner similar to the case of nlYM$_2$.

\section{nlYM$_2$: definition and the classical properties}
We define the nlYM$_2$ with the following action.
\be\label{3}
e^S:=\int\rD B\;\exp\left\{\int\rd\mu\;i{\rm tr}(BF)+w\left[\int\rd\mu\;
{\rm tr}(B^2)\right]\right\}.
\ee
Here $\rd\mu$ is the invariant measure of the surface:
\be
\rd\mu:={1\over 2}\epsilon_{\mu\nu}\rd x^\mu\rd x^\nu,
\ee
$B$ is a pseudo--scalar field in the adjoint representation of the group,
and $F$ is the field strength corresponding to the gauge field.
The classical equation for the gauge field is
\be\label{4}
{{\delta S}\over{\delta A}}=0,
\ee
or equivalently,
\be\label{5}
{{\delta}\over{\delta A}}e^S=0.
\ee
To obtain this, we note that
\be\label{6}
{{\delta}\over{\delta F}}e^S=i\int\rD B\;B\exp\left\{\int\rd\mu\;i{\rm tr}
(BF)+w\left[\int\rd\mu\;{\rm tr}(B^2)\right]\right\}.
\ee
The integral in the right--hand side is not easy to be carried out. But,
using the fact that the only tensor appearing in the right--hand side is
the {\it metric}
\be\label{7}
{\cal G}_{ab}(x,y):=\omega_{ab}\delta(x,y),
\ee
where $\omega$ is the metric defined on the algebra, it is seen that
there exists a function ${\cal A}$ such that
\be\label{8}
{{\delta}\over{\delta F}}e^S=F{\cal A}\left[\int\rd\mu\;{\rm tr}(F^2)
\right],
\ee
or in a more explicit form,
\be\label{9}
{{\delta}\over{\delta F^a(x)}}e^S=F_a(x){\cal A}\left[\int\rd\mu\;{\rm tr}
(F^2)\right].
\ee
From this, one can obtain the left--hand side of (\ref{5}) as
\bea\label{10}
{{\delta}\over{\delta A^a(x)}}e^S&=&\int\rd\mu'{{\delta}\over{\delta
F^b(x')}}e^S{{\delta F^b(x')}\over{\delta A^a(x)}}\cr
&=&{\cal A}\left[\int\rd\mu\;{\rm tr}(F^2)\right] \int\rd\mu'\;F_b(x')
{{\delta F^b(x')}\over{\delta A^a(x)}}.
\eea
Comparing this to the variation of the YM action, it is seen that
\be\label{11}
{{\delta}\over{\delta A^a(x)}}e^S=-2{\cal A}\left[\int\rd\mu\;{\rm tr}(F^2)
\right]{{\delta}\over{\delta A^a(x)}}S^{\rm YM}.
\ee
This shows that classical solutions of the YM theory are also classical
solutions of the nlYM theory.

\section{The partition function and the wave functions of nlYM's}
Along the lines of \cite{BlTh, KhAl}, we begin by calculating the
wave--function of a disk. We have
\bea\label{12}
\psi_D(U)&=&\int\rD F\; e^S\delta\left(\Pexp\oint_{\partial D}A, U\right)
\cr
&=&\int\rD B\;\rD F\exp\left\{\int\rd\mu\;i{\rm tr}(BF)+w\left[\int\rd\mu\;
{\rm tr}(B^2)\right]\right\}\cr
&&\times\delta\left(\Pexp\oint_{\partial D}A, U\right).
\eea
Here $U$ is the class of the Wilson loop corresponding to the boundary. The
delta function is also a class delta function, that is, its support is where
its two arguments are in the same congugacy class. This delta function can
be expanded in terms of the characters of irreducible unitary
representations of the group:
\be\label{13}
\delta\left(\Pexp\oint_{\partial D}A, U\right)=\sum_R\chi_R(U^{-1})\chi_R
\left(\Pexp\oint_{\partial D}A\right).
\ee
One can now introduce Fermionic variables $\eta$ and $\bar\eta$ in the
representation $R$ to write the Wilson loop as \cite{KhAl}
\be\label{14}
\chi_R\left(\Pexp\oint_{\partial D}A\right)=\int\rD\eta\;\rD\bar\eta
\left[\int_0^1\rd t\;\bar\eta (t)\dot\eta (t)+\oint_{\partial D}\bar\eta
A\eta\right]\eta^\alpha(0)\bar\eta_\alpha(1).
\ee
Inserting (\ref{14}) in (\ref{13}) and then (\ref{12}), using the
Schwinger--Fock gauge, and integrating over, $F$, $B$, and the fermionic
variables, respectively, one arrives at
\be\label{15}
\psi_D(U)=\sum_R\chi_R(U^{-1})d_R\exp\{ w[-AC_2(R)]\},
\ee
where $d_R$ is the dimension of the representation $R$ and $C_2(R)$ is the
second Casimir of the representation $R$.
The details of calculation are the same as those done in \cite{KhAl}.
Note, however, that one cannot simply glue the disk wave--functions to
obtain, for example, the wave function corresponding to a larger disk or
that of a sphere. The reason is that the action of the original $B$--$F$
theory is not extensive, that is
\be\label{16}
S_{A_1+A_2}(B, F)\ne S_{A_1}(B, F)+S_{A_2}(B, F).
\ee
So, if the disk $D$ is divided into two smaller disks $D_1$ and $D_2$,
\be\label{17}
\psi_{D}(U)\ne\int\rd U_1\;\psi_{D_1}(UU_1)\psi_{D_2}(UU_1^{-1}).
\ee
Two obtain the wave function for an arbitrary surface, however, one can
begin with a disk of the same area and impose boundary conditions on certain
parts of the boundary of the disk. These conditions are those corresponding
to the identifications needed for constructing the desired surface from a
disk. This is done in exactly in the same manner as the case of YM$_2$ or
gYM$_2$, as the only things to be calculated are integrations over group of
characters of the same representation \cite{ZPhys}. This is easily done and
one arrives at
\bea\label{18}
\psi_{\Sigma_{g,q}}(U_1,\cdots,U_n)&=&\sum_R h_R^{q}d_R^{2-2g-q-n}\cr
&&\times\chi_R(U_1^{-1})\cdots\chi_R(U_n^{-1})\exp\{w[-C_2(R)
A(\Sigma_{g,q})]\}.
\eea
Here $\Sigma_{g,q}$ is a surface containing $g$ handles and $q$ projective
planes. It has also $n$ boundaries. This is the most general surface with
finite area. $h_R$ is defined through
\be\label{19}
h_R:=\int\rd U\chi_R(U^2),
\ee
and it is zero unless the representation $R$ is self--conjugate. In this
case, this representation has an invariant bilinear form. Then, $h_R=1$ if
this form is symmetric and $h_R=-1$ if it is antisymmetric \cite{BrDi}.

As a special result, the partition function of the theory on a sphere is
obtained if we put $U_i$'s equal to unity and $q$ and $g$ equal to zero.
We arrive at
\be\label{20}
Z=\sum_Rd_R^2\exp\{w[-AC_2(R)]\}.
\ee

\section{Large--$N$ limit of nlYM$_2$}
Starting from (\ref{20}), consider the case the gauge group is $U(N)$.
The representations of this group are labeled by $N$ integers $n_i$
satisfying
\be\label{21}
n_i\geq n_j,\qquad i<j.
\ee
The dimension of this representation is
\be\label{23}
d_R=\prod_{1\leq i<j\leq N}\left(1+{{n_i-n_j}\over{j-i}}\right),
\ee
and the $l$-th Casimir is
\be\label{24}
C_l(R)=\sum_{i=1}^N[(n_i+N-i)^l-(N-i)^l].
\ee
For $l=2$, one can redefine the function $w$ and introduce another function:
\be\label{25}
w(-AC_2)=:-N^2W(A\tilde C_2),
\ee
where
\be\label{26}
\tilde C_l(R):={1\over{N^{l+1}}}\sum_{i=1}^N(n_i+N-i)^l.
\ee
Then, following \cite{Rus}, we use the definitions
\be\label{27}
x:=i/N,
\ee
and
\be\label{28}
\phi:={{i-n_i-N}\over{N}},
\ee
to write the partition function as
\be\label{29}
Z=\int\rD\phi\;\exp[S(\phi)],
\ee
where
\be\label{30}
S(\phi):=-N^2\left\{W\left[A\int_0^1\rd x\;\phi^2(x)\right]+\int_0^1\rd x
\int_0^1\rd y\;\log|\phi(x)-\phi(y)|\right\}.
\ee
In the large--$N$ limit, only the representation (i.e. the configuration of
$\phi$) contributes to the partition function that maximizes $S$. To find
it, one puts the variation of $S$ with respect to $\phi$ equal to zero:
\be\label{31}
-A\; W'\left[A\int\rd y\;\phi^2(y)\right]\times 2\phi(x)+2\rP\int{{\rd y}
\over{\phi(x)-\phi(y)}}=0.
\ee
Defining
\be\label{32}
\tilde A:=A\;W'\left[A\int\rd y\;\phi^2(y)\right],
\ee
the equation for $\phi$ is written as
\be\label{33}
{{\tilde A}\over 2}[2\phi(x)]=\rP\int{{\rd y}\over{\phi(x)-\phi(y)}}.
\ee
This equation is the same as that obtained in \cite{Rus} and \cite{DoKa},
and can be solved in the same manner. First, one defines a density function
for $\phi$ as
\be\label{34}
\rho(z):={{\rd x}\over{\rd\phi(x)}}\Big|_{\phi(x)=z}.
\ee
Then, (\ref{33}) becomes
\be\label{35}
{{\tilde A}\over 2}(2z)=\rP\int_{-a}^a{{\rd w\;\rho(w)}\over{z-w}}.
\ee
Next, a function $H$ is defined on the complex plane through
\be\label{36}
H(z)=\int_{-a}^a{{\rd w\;\rho(w)}\over{z-w}}.
\ee
Here, $z$ is a complex variable. This function is analytic on the complex
plane, except for $z\in[-a,a]$, where $H$ has a branch cut. Also, from the
definition of $\rho$, it is seen that
\be\label{37}
\int_{-a}^a\rd w\;\rho(w)=1,
\ee
which shows
\be\label{38}
H(z)\sim{1\over z},\qquad z\to\infty.
\ee
The function $H$ is then calculated to be \cite{DoKa}
\be\label{39}
H(z)={1\over{2\pi i}}\sqrt{z^2-a^2}\oint_c{{(\tilde A/2)(2\lambda)
\rd\lambda}\over{(z-\lambda)\sqrt{\lambda^2-a^2}}},
\ee
where the integration contour encircles the cut $[-a,a]$ but the point $z$
is outside it. The integration is readily done and one arrives at
\be\label{40}
H(z)=\tilde A(z-\sqrt{z^2-a^2}).
\ee
To obtain $a$, one can use (\ref{38}), which yields
\be\label{41}
a=\sqrt{{2\over{\tilde A}}}.
\ee
From (\ref{36}), it is seen that
\be\label{42}
-\pi\rho(z)={\rm Im}H(z+i\epsilon),\qquad \hbox{for {\it z} real},
\ee
which gives
\be\label{43}
\rho(z)={{\tilde A}\over\pi}\sqrt{{2\over{\tilde A}}-z^2}.
\ee
This is, of course, in complete accordance with \cite{Rus,DoKa}. But one
must now obtain the density in terms of $A$ not $\tilde A$. To do so, let
us return to the definition (\ref{32}). First, we need the integral
\be\label{44}
\int\rd x\;\phi^2(x)=\int\rd z\;\rho(z)z^2,
\ee
which is the coefficient of $1/z^3$ in the large--$z$ expansion of $H$. This
is calculated to be
\bea\label{45}
\int\rd z\;\rho(z)z^2&=&{{\tilde A a^4}\over 8}\cr
                     &=&{1\over{2\tilde A}}.
\eea
From this, one obtains an equation for $\tilde A$ as
\be\label{46}
\tilde A=A\;W'\left({A\over{2\tilde A}}\right),
\ee
or
\be\label{47}
{{Aa^2}\over 4}W'\left({{Aa^2}\over 4}\right)={1\over 2}.
\ee

Defining a free--energy function as
\be\label{48}
f:=-{1\over{N^2}}S\Big|_{\phi_{\rm cla.}},
\ee
It is seen that
\bea\label{49}
f'(A)&=&\int\rd x\;\phi^2(x)W'\left[A\int\rd y\;\phi^2(y)\right]\cr
     &=&{{\tilde A}\over A}\int\rd x\;\phi^2(x).
\eea
or
\be\label{49-1}
f'(A)={1\over{2A}}.
\ee
The function $W$ is disappeared from $f'$, as it can be seen by the
rescaling $\tilde\phi:=\sqrt{A}\phi$.

This completes our discussion of the weak--region nlYM. As $A$ increases,
a situation is encountered where $\rho$ exceeds 1. This density function is,
however, not acceptable, as it violates the condition (\ref{21}). The value
of $A$ at which this occurs is obtained from
\be\label{50}
\rho(0)=1,
\ee
which gives
\be\label{51}
\tilde A_c={{\pi^2}\over 2}.
\ee
For $\tilde A>\tilde A_c$, one must take an ansatz for $\rho$ as
\be\label{52}
\rho_s(z)=\cases{\tilde\rho_s(z),&$z\in L$\cr
                 1,&$z\in [-b,b]$\cr},
\ee
where
\be\label{53}
L:=[-a,-b]\cup[a,b]
\ee
Using methods exactly the same as those used in \cite{DoKa,AgAlKh,AlKhAg},
one must solve
\be\label{54}
{{\tilde A}\over 2}(2z)=\rP\int_{-a}^a{{\rd w\;\rho_s(w)}\over{z-w}},
\qquad z\in L.
\ee
To do so, one defines a function $H_s$ as
\be\label{55}
H_s(z)=\int_{-a}^a{{\rd w\;\rho_s(w)}\over{z-w}},
\ee
which is found to be
\bea\label{56}
H_s(z)&=&\log{{z+b}\over{z-b}}+{{\sqrt{(z^2-a^2)(z^2-b^2)}}\over{2\pi i}}\cr
       &&\times\oint_{c_L}\rd\lambda{{(\tilde A/2)(2\lambda)-
       \log[(\lambda +b)/(\lambda -b)]}\over{(z-\lambda)\sqrt{
       (\lambda^2-a^2)(\lambda^2-b^2)}}}\cr
      &=&{{\tilde A}\over 2}(2z)-\sqrt{(z^2-a^2)(z^2-b^2)}\cr
      &&\times\int_{-b}^b{{\rd\lambda}\over{(z-\lambda)
      \sqrt{(a^2-\lambda^2)(\lambda^2-b^2)}}}.
\eea
Here $c_L$ is a contour encircling $L$, leaving $[-b,b]$ and the point $z$
out. Note that everything is exactly the same as the case of ordinary
YM theory, except that $A$ is replaced by $\tilde A$. Using the fact that
$H_s$ should behave as $1/z$ for large $z$, one obtains two equations
\be\label{58}
\tilde A=\int_{-b}^b{{\rd\lambda}\over{\sqrt{(a^2-\lambda^2)
(\lambda^2-b^2)}}}
\ee
and
\be\label{59}
1=\int_0^b\rd\lambda{{a^2+b^2-\lambda^2}\over{\sqrt{(a^2-\lambda^2)
(\lambda^2-b^2)}}}.
\ee
For $\tilde A$ near $\tilde A_c$, these equations are solved as
\be\label{60}
{1\over a}={\pi\over 2}\left(1+{{\pi^2b^2}\over{16}}+{{\pi^4b^4}\over{128}}
\right)
\ee
and
\be\label{61}
\tilde A={{\pi^2}\over 2}\left(1+{{\pi^2b^2}\over 8}+{{7\pi^4b^4}\over{256}}
\right).
\ee
Now, using (\ref{49}), and the fact that the integral $\int\rd x\;\phi^2(x)$
is the coefficient of $1/z^3$ in the large--$z$ expansion of $H_s$, one
arrives at
\be\label{62}
f_s'(A)={{\tilde A}\over A}\left[{1\over{2\tilde A}}+
{{(\tilde A-\tilde A_c)^2}\over{\tilde A_c^3}}\right]+O[
(\tilde A-\tilde A_c)^3],
\ee
or
\bea\label{63}
f_s'-f_w'&=&{{(\tilde A-\tilde A_c)^2}\over{A_c\tilde A_c^2}}\cr
         &=&\left({{\rd\tilde A}\over{\rd A}}\right)^2_{c,s}{{(A-A_c)^2}
         \over{A_c\tilde A_c^2}}.
\eea
This shows that the third--order phase transition is there, unless the
derivative of $\tilde A$ with respect to $A$ is zero at the critical point
in the strong region.

\section{nlgYM$_2$: wave functions, partition function, and the large--$N$
limit}
A nonlocal generalized Yang--Mills theory is defined by the action
\be\label{64}
e^S:=\int\rD B\;\exp\left\{\int\rd\mu\;i{\rm tr}(BF)+w\left[\int\rd\mu\;
\Lambda(B)\right]\right\}.
\ee
Following \cite{KhAl}, and using the same technique of section 3, one can
find the wave function on a disk:
\be\label{65}
\psi_D(U)=\sum_R\chi_R(U^{-1})d_R\exp\{ w[AC_\Lambda(R)]\},
\ee
where
\be\label{66}
C_\Lambda(R)1_R:=\Lambda(-iT_R),
\ee
and $\Lambda(-iT_R)$ means that one has put $-iT^a$ in the representation
$R$ instead of $B^a$ in the function $\Lambda$. Using the same technique of
section 3, one can construct the wave function on an arbitrary surface as
\bea\label{67}
\psi_{\Sigma_{g,q}}(U_1,\cdots,U_n)&=&\sum_R h_R^{q}d_R^{2-2g-q-n}\cr
&&\times\chi_R(U_1^{-1})\cdots\chi_R(U_n^{-1})\exp\{w[C_\Lambda(R)
A(\Sigma_{g,q})]\},
\eea
and the partition function on the sphere as
\be\label{68}
Z=\sum_Rd_R^2\exp\{w[AC_\Lambda(R)]\}.
\ee

The large--$N$ limit of this theory (for the gauge group $U(N)$), is defined
by suitable rescalings. Taking $C_\Lambda$ a linear function of the
rescaled Casimirs (\ref{26}), one can define a function $W$ as
\be\label{69}
-N^2W\left[A\sum_l \alpha_l\tilde C_l(R)\right]:=w[AC_\Lambda(R)].
\ee
In the large--$N$ limit, the partition function becomes like (\ref{29}),
but with
\be\label{70}
S(\phi):=-N^2\left\{W\left[A\int_0^1\rd x\;G(\phi)\right]+\int_0^1\rd x
\int_0^1\rd y\;\log|\phi(x)-\phi(y)|\right\},
\ee
where
\be\label{72}
G(\phi):=\sum_l(-1)^la_l\phi^l.
\ee
Then, following the procedure of section 4, that representation contributes
to the partition function that maximizes $S$. The equation for this
representation is
\be\label{72-1}
-A\; W'\left[A\int\rd y\;G(\phi)\right]G'[\phi(x)]+2\rP\int{{\rd y}\over
{\phi(x)-\phi(y)}}=0.
\ee
Defining
\bea\label{73}
\tilde A&:=&A\;W'\left\{A\int\rd y\;G[\phi(y)]\right\}\cr
         &=&A\;W'\left[A\int\rd z\;\rho(z)\;G(z)\right],
\eea
the equation for $\phi$ is written as
\be\label{74}
{{\tilde A}\over 2}G'[\phi(x)]=\rP\int{{\rd y}\over{\phi(x)-\phi(y)}},
\ee
or, in terms of the density $\rho$,
\be\label{75}
{{\tilde A}\over 2}G'(z)=\rP\int{{\rd w\;\rho(w)}\over{z-w}}.
\ee
Note that equation (\ref{75}) does not contain $W$. That is, the solution
for $\rho$ is the same as that obtained for gYM$_2$. The only difference is
that one should put $\tilde A$ instead of $A$. This is true for the weak
region, as well as for the strong region. One can also obtain the derivative
of the free energy with respect to $A$, in the same way it was obtained in
section 4. We have
\bea\label{76}
f'(A)&=&\int\rd x\;G[\phi(x)]W'\left\{A\int\rd y\;G[\phi(y)]\right\}\cr
     &=&{{\tilde A}\over A}\int\rd z\;\rho(z)\;G(z).
\eea
The set of equations to be solved are (\ref{75}) and (\ref{37}), to obtain
$\rho$ in terms of $\tilde A$; then (\ref{73}) to obtain $\tilde A$ in terms
of $A$. Also note that $f'(A)$ is essentially the same as $f'(A)$ for the
corresponding {\it local} gYM$_2$. The differences are the existence of
$\tilde A$ instead of $A$, and an overall factor $\tilde A/A$. The
similarity between the nonlocal theory and the local theory holds also in
the strong region. Since there, one still has equation (\ref{75}) for the
set where $\rho<1$. That is, here too the equations for $\rho$ are the same
as those of the corresponding local theory.


\newpage
\begin{thebibliography}{99}
\bibitem{BlTh} M. Blau \& G. Thomson; ``Lectures on 2d Gauge Theories,
               Proceedings of the 1993 Trieste Summer School on High Energy
               Physics and Cosmology'' (World Scientific, Singapore, 1994)
               175.
\bibitem{Wit} Edward Witten; J. Geom. Phys. {\bf 9} (1992) 303.
\bibitem{GaSoYa} O. Ganor, J. Sonnenschein, \& S. Yankelowicz; Nucl. Phys.
                 {\bf B434} (1995) 139.
\bibitem{KhAl} M. Khorrami \& M. Alimohammadi; Mod. Phys. Lett. {\bf A12}
               (1997) 2265.
\bibitem{Rus} B. Rusakov; Phys. Lett. {\bf B303} (1993) 95.
\bibitem{DoKa} M. R. Douglas \& V. A. Kazakov; Phys. Lett. {\bf B319} (1993)
               219.
\bibitem{AgAlKh} A. Aghamohammadi, M. Alimohammadi, \& M. Khorrami; Mod.
                 Phys. Lett. {\bf A14} (1999) 751.
\bibitem{AlKhAg} M. Alimohammadi, M. Khorrami, \& A. Aghamohammadi; Nucl.
                 Phys. {\bf B510} (1998) 313.
\bibitem{AlTo} M. Alimohammadi \& A. Tofighi; Eur. Phys. J. {\bf 8} (1999)
               711.

\bibitem{ZPhys} M. Alimohammadi \& M. Khorrami; Z. Phys. {\bf C76} (1997)
                729.
\bibitem{BrDi} T. Brocker \& T. T. Dieck; ``Representations of Compact Lie
               Groups'' (Springer, 1985).
\end{thebibliography}
\end{document}